\documentclass[prl,twocolumn]{revtex4}
\usepackage{epsfig}
\usepackage{amsmath}
\usepackage{epsfig}

\begin{document}
\title
{Expectation values of $p^2$ and $p^4$ in the square well potential}    
\author{Zafar Ahmed$^1$, Dona Ghosh$^2$, Sachin Kumar$^3$, Joseph Amal Nathan$^4$}
\affiliation{$^1$Nuclear Physics Division, $^3$Theoretical Physics Section, $^4$Reactor Physics Design Division, Bhabha Atomic Research Centre, Mumbai 400 085, India\\
$^2$Department of Mathematics, \nopagebreak Jadavpur University. Jadavpur, Kolkata, 700032, India}
\email{1:zahmed@barc.gov.in, 2:rimidonaghosh@gmail.com,  3:sachinv@barc.gov.in, 4:josephan@barc.gov.in}
\date{\today}
\begin{abstract}
Position and momentum representations of a wavefunction	
$\psi(x)$ and $\phi(p)$, respectively are physically equivalent yet mathematically in a given case one may be easier or more transparent than the other. This disparity may be so much so that one has to device a special strategy to get the quantity of interest in one of them. We revisit finite square well (FSW) in this regard. Circumventing the the problems of discontinuity of second and higher derivatives of $\psi(x)$ we obtain simple analytic expressions of $<\!p^2\!>$ and $<\!p^4\!>$. But it is the surprising fall-off of $\phi(p)$ as $p^{-6}$ that reveals and restricts $<\!p^s\!>$ to be finite and non-zero only for $s=2,4$. In finding $<\!p^s\!>(s=2,4)$ from $\phi(p)$, $p$-integrals are improper which for time-being, have been evaluated numerically to show the agreement between two representations. 
\end{abstract}
\maketitle
\section{I. Introduction}
In the microscopic world the crucial elementary part of a system of mass $m$ is taken to be in a potential $V(x)$ and the state $\psi(x)$ of the system in position space is governed by the Schor{\"o}dinger equation [1]
\begin{equation}
\frac{d^2\psi(x)}{dx^2}+\frac{2m}{\hbar^2}[E-V(x)] \psi(x)=0.
\end{equation}
Solutions of most of the problems of quantum mechanics have been found by solving (1) analytically or numerically. This equation can be written in momentum space by the Fourier transformation 
${\cal F}$$[\psi(x)]$ 
\begin{equation}
\phi(p)={\cal F}[\psi(x)]=(2\pi \hbar)^{-1/2} \int_{-\infty}^{\infty} \psi(x) e^{-ipx/\hbar} dx,
\end{equation}
of the differential equation (1) to an integral equation as
\begin{equation}
[p^2/2m -E] \phi(p)= \int_{-\infty}^{\infty} U(p-p') \phi(p') dp'.
\end{equation}
$U(p)$=${\cal F}$$[V(x)]$. Textbooks [1] emphasize the physical equivalence of  these two representations. Solving integral equation for even solvable potentials of position space  is usually difficult. One may 
solve (3) for the bound state problem of $V(x)=-\lambda \delta(x)$ which is simple and quick [2]. Interestingly, for harmonic oscillator  the integral equation gets changed to exactly the same equation as (1) with $x$ changed 
by  $p$ [3]. Morse oscillator has also been solved by changing (3) into a differential equation [3]. The other way to get $\phi(p)$ for a potential model $V(x)$ is by finding the Fourier transform (2) of the corresponding $\psi(x)$. An interesting collection of $\phi(p)$ for various models is available in Ref. [4,5].

Apart from minor exceptions, for a given potential $V(x)$ and for a fixed query, one of $\psi(x)$ and $\phi(p)$ may be simple or revealing but the other one may be difficult or un-revealing, so much so that special strategies may be required to find it and the quantities thereof. Minor exceptions where the two approaches are identical/similar are e.g., the Fourier transforms of two ground states $\psi_0(x)= A~e^{-x^2/2}$ and $B~\mbox{sech} (x/2)$  are $\phi_0(p)=A~e^{-p^2/2}$ and $\sqrt{2\pi}B~\mbox{sech}p \pi$, respectively. Normally, 
these two representations yield different mathematical functions. For instance, for the Dirac delta potential well $V(x)=-(\lambda \hbar^2/m) \delta(x)$, $\psi(x)= e^{-\lambda|x|}$ (non-differentiable at $x=0$) but the corresponding $\phi(p)=\sqrt{2/\pi}\lambda/(\lambda^2+p^2)$ (infinitely differentiable) shows rather clearly that $<\!p^2\!>=$finite but $<\!p^4\!>=\infty$. Moreover, the calculation of $<\!p^2\!>$ in position space requires carefulness [6]. In another instance, note that infinite square well (ISW) $<\!p^2\!>$ can be obtained easily in position space but we point out that in momentum space the integral becomes improper [7] which may diverge unless we are careful (see section II below).

Textbooks [1] of quantum mechanics discuss (infinite) finite square well (FSW) universally. Both FSW and ISW are also called box potentials, Cummings [8] pointed out that particle in a box is not simple. According to him, square well which is  discontinuous  at the end points $x=\pm a$ has the second derivative of eigenfunctions namely $\frac{d^2\psi(x)}{dx^2}$ as  discontinuous. Cummings showed that calculation of expectation values of even  powers of momentum $p^s$ $(s=2,4,...)$ using position space eigenfunctions $\psi(x)$ poses problems of discontinuities  in terms of Dirac delta $\delta(x\pm a)$ and its derivatives. He argued that for such a well $<\!p^s\!>$ is finite for $s=2,4$ but infinite for  $s=6,8,...$. However, he did not find the analytic forms of $<\!p^s\!> (s=2,4)$ though his analysis was confined to square well. It may be pointed out that  expectation values of odd powers of $p$  vanish due to anti-symmetry of integrands. Particularly, for ISW  he showed [8] that only $<\!p^2\!>$ is finite and $<\!p^s\!>$ $(s=4,6,...)$ is infinite. Here, we show that in the ISW potential,  $<\!p^2\!>$ in momentum space may diverge unless we realize that it is improper but convergent [7].

Circumventing the problem of discontinuities in various derivatives of $\psi(x)$. we find  simple expressions of $<\!p^2\!>$ and $<\!p^4\!>$ for finite square well (FSW) potential.
The other way of obtaining $<\!p^2\!>$ or $<\!p^4\!>$ is to find the momentum space eigenfunctions $\phi(p)$ from the Fourier transform (1) of $\psi(x)$ then one can evaluate second and fourth moments of the distribution function $I(p)= |\phi(p)|^2$. For the finite square well potential  $I(p)$
has been shown to fall-off asymptotically as $p^{-6}$ [8]. The acclaimed fall-off ($p^{-6})$ is both correct and surprising but the details of the analytic forms of $\phi(p)$ in [8] are unfortunately incorrect in several ways. Apparently, interesting papers [4,5] on momentum distribution of particle in one-dimensional potentials seem to have left out this interesting model of FSW for this paper [1]. In this paper, we propose to re-derive $\phi(p)$ for FSW, we shall compare our results with those obtained by numerical integrations for a  confirmation. We shall be using the same parametrization as adopted in [8] for a ready and convenient comparison.

\section{ II. Infinite square well (ISW) potential}
Infinite square well potential [1,8] is written as\\
\begin{equation}
V(|x| < a/2) = 0,\quad V(|x|\ge a/2)=\infty.
\end{equation}
The even and odd parity bound state  eigenfunctions of (1) are well known for $(-a/2 \le x \le a/2)$ respectively as 
\begin{eqnarray}
\psi_n (x)= \sqrt{2/a}~ \cos\beta_n x,\quad  (n- \mbox{odd})\nonumber \\ \psi_n(x)= \sqrt{2/a}~ \sin \beta_n x, \quad (n-\mbox{even}).
\end{eqnarray}
Here $\beta_n=\sqrt{2mE_n/\hbar^2}$, $E_n=n^2\hbar^2/(2ma^2)$.
One can readily verify that
\begin{equation}
<\psi_n(x)|p^2|\psi_n(x)>=\beta_n^2.
\end{equation} 
Similarly, one can also find $<\!p^4\!>=\beta_n^4$ which would actually contradict when we try to get it from the corresponding $\phi(p)$ obtained from (2) and these are known as [8]
\begin{eqnarray}
\phi_n(p)=N_n \cos \left (\frac{pa}{2\hbar}\right)/(n^2\pi^2\hbar^2-p^2a^2),\quad (n-\mbox{odd}) \nonumber \\
\phi_n(p)=N_n \sin \left (\frac{pa}{2\hbar}\right)/(n^2\pi^2\hbar^2-p^2a^2), \quad (n-\mbox{even}).
\end{eqnarray}
Here $N_n=\sqrt{4a n^2 \pi\hbar^3} (-1)^{(n+1)/2}$. Notice that $I(p)=|\phi(p)|^2$.  In both cases  (7)  $I(p)$ appears to diverge when $p=n\pi \hbar/a$, but more carefully one can see that $I(n\pi \hbar/a)=0/0$, using L'Hospital rule we get $\lim_{p\rightarrow n \pi \hbar/a} I(p) = a/(4 \pi \hbar   )$ (finite). Therefore the integral in finding $<p^2>$ using (7) is improper [7] but convergent and one can recover (6).  Next, one can see that for (7), $I(p) \sim p^{-4}$, consequently the integrals  evaluating $<p^s>$ using Eq. (7) will diverge for $s=4,6,8,..$,
a fact which is not revealed by the position space eigenfunctions in (5) for ISW. In Ref. [8], the divergence of $<p^4>$  has been predicted and attributed to the rigid walls at $x=\pm a/2$ in ISW (4), at these points there exist discontinuities of Dirac delta function and its derivatives in $p^4 \psi(x)$. We  discus FSW in  position space in the next section.

\section{III. Finite square well potential in position space}
The finite square well (FSW) model is written as
\begin{equation}
V(|x|\ge a/2)= V_0, \quad V(|x|<a/2)=0,
\end{equation}
which is discussed universally in  textbooks [1] of quantum mechanics.
Let us introduce the following definitions as per the Ref. [8]
\begin{equation}\alpha= \sqrt{\frac{2m(V_0-E)}{\hbar^2}}, \beta=\sqrt{\frac{2m E}{\hbar^2}}, d=\frac{\beta a}{2}, z=\frac{p}{\hbar}.
\end{equation}
The solution of Schr{\"o}dinger equation (1) for even parity is
\begin{eqnarray}
\psi_e(|x|\le a/2)=A_e \cos(\beta x),\nonumber \\  \psi_e(|x|>a/2)= A_e\cos d~ \exp[-\alpha(|x|-a/2)],
\end{eqnarray}
with eigenvalue equation as 
\begin{equation}
\tan (\beta a/2)= \alpha/\beta.
\end{equation}
The odd parity solution is
\begin{eqnarray}
\psi_o(|x|\le a/2)=A_o \sin(\beta x),\nonumber \\  \psi_o(|x|>a/2)= A_o ~\mbox{sgn}(x)\sin d~ \exp[-\alpha(|x|-a/2)],
\end{eqnarray}
with eigenvalue equation as
\begin{equation}
\tan (\beta a/2) =-\beta/\alpha.
\end{equation}
Here $\mbox{sgn}(x)=-1$, if $x<0$ and 1, if $x>0$. In this model $\psi(x)$ are easily normalizable,
the correct forms for the normalization coefficients $A_e, A_o$ are 
\begin{eqnarray}
A_e=2^{1/2}[[1+\cos(2d)]/\alpha+\sin(2d)/\beta+a]^{-1/2},\\ \nonumber 
A_o=  2^{1/2}[[1-\cos(2d)]/\alpha-\sin(2d)/\beta+a]^{-1/2}.
\end{eqnarray}
Compare the above equations with Eqs. A.3 and A.5 of Ref.[8],  where expressions of $A_e$ and $A_o$ have got interchanged inadvertently. 

Next, we propose to find $<\!\psi(x)|p^2\psi(x)\!>$  by using 
the Schr{\"o}dinger equation (1) itself as $p^2\psi(x)=(2m/\hbar^2)[E-V(x)]$ for a  simple way of circumventing the discontinuity of $\frac{d^2\psi}{dx^2}$ at $x=\pm a/2$. For FSW by using $\psi(x)$ (10,12), we obtain
\begin{eqnarray}
<\! \psi_e(x)|\frac{2m}{\hbar^2}[E{-}V(x)]|\psi_e(x)\!>=\beta^2{-}\frac{2mV_0 A^2_e \cos^2 d}{\alpha \hbar^2}, \nonumber \\ 
<\! \psi_o(x)|\frac{2m}{\hbar^2}[E{-}V(x)]|\psi_o(x)\!>{=}\beta^2{-}\frac{2mV_0 A^2_o \sin^2 d}{\alpha \hbar^2}.
\end{eqnarray}
Notice that $V(x)$ has (finite) jump discontinuity at $x=\pm a$.
Here we have also utilized the fact that in a definite integral discontinuity of finite jump at a finite number of points of the integrand is allowed (piece-wise integration).
When we take limit $V_0\rightarrow \infty$ in the above equations, the second terms vanish and  we recover the result (6) of ISW. The expectation value of $p^2$ for square well potential seems to 
a common question of students where, unwary may be found in an embarassing situation. 
Further, for finding $<\!p^4\!>= <\psi(x)|p^4|\psi(x)>$, we utilize the Hermiticity of $p^2$ to write  it as $<p^2\psi(x)|p^2\psi(x)>$ and get 
\begin{equation}
<\!\psi(x)|p^4|\psi(x)\!>{=}{\left(\frac{2m}{\hbar}\right)^2}<\! \psi(x)|[E{-}V(x)]^2|\psi(x)\!>.
\end{equation}
By using the eigenfunctions (10,12) of FSW in above, we obtain
\begin{eqnarray}
<\!p^4\!>=\beta^4+\left(\frac{2m}{\hbar}\right)^2 \frac{A^2_e}{\alpha} [V_0^2-2V_0\beta^2]\cos^2d, \nonumber \\
<\!p^4\!>=\beta^4+\left(\frac{2m}{\hbar}\right)^2 \frac{A^2_o}{\alpha} [V_0^2-2V_0\beta^2]\sin^2d.
\end{eqnarray}
Here, it can be readily checked that in limit $V_0\rightarrow \infty$ the expressions above become $\infty$, confirming that for ISW $<p^4>$ diverges [8]. 

\section{ IV. Finite square well potential in momentum space}
An interesting feature of the Fourier transform is that if $\psi(x)$ is normalized so is $\phi(p)$.
Here, we shall be using the normalized eigenfunctions of FSW given in Eqs. (10) and (12).
The Fourier transform  $\phi(p)$ of $\psi(x)$ (10,12) in (2) will have two parts $\phi^{in}(p)$ for $|x| \le a/2$ and $\phi^{out}(p)$ for $|x| > a/2$, we find them as
\begin{multline}
\hspace*{-0.4cm}\phi^{in}_e(p){=}\frac{2 A_e [\beta \cos(za/2) \sin d{-}z \sin(za/2) \cos d]}{\sqrt{2\pi \hbar}(\beta^2{-}z^2)}, \\ 
\hspace*{-0.4cm}\phi^{out}_e(p){=}\frac{2 A_e [\alpha \cos(za/2){-}z \sin(za/2)]\cos d}{\sqrt{2\pi \hbar} (z^2{+}\alpha^2)}, z{=}\frac{p}{\hbar}.
\end{multline}
In above, apparently $\phi^{in}_e(p)$ and $\phi^{out}_e(p) $ vary asymptotically as $p^{-1}$, giving the impression that even $<\!p^2\!>$ would diverge. However, most interestingly, by using the eigenvalue condition (11), $\phi_e(p)=\phi^{in}_e(p)+\phi^{out}_e(p)$ simplifies to 
the inspiring form given as
\begin{eqnarray}
\phi_e(p){=} \frac{2 A_e \gamma^2 [-\alpha \cos(za/2)+ z \sin(za/2)] \cos d}{\sqrt{2\pi \hbar} (z^2+\alpha^2)(z^2-\beta^2)}   {\sim}\frac{1}{p^{3}}, \hspace*{.2 cm}
\end{eqnarray}
here $\gamma^2=\alpha^2+\beta^2=2mV_0/\hbar^2$.
It can be seen that $I_e(p)=|\phi_e(p)|^2 \sim p^{-6}$ as in Eq. A.4 of [8], however the  details of (19) show  disagreements.
$\phi_e(\hbar \beta)$ in [8] diverges, but in our result (19) $\phi_e(\hbar \beta)=0/0$, next by finding the limit by L'Hospital's rule  we find that
\begin{eqnarray}
\hspace*{-.15 cm}\lim_{z\rightarrow \beta} \frac{[{-}\alpha \cos(\frac{za}{2})+z \sin(\frac{za}{2})]}{(z^2-\beta^2)} {\rightarrow} 
\frac{(\alpha a{+}2)\sin d{-}2d \cos d}{4\beta},
\end{eqnarray}
which is finite. 

\begin{figure}
	\centering
	\includegraphics[width=8 cm,height=5 cm]{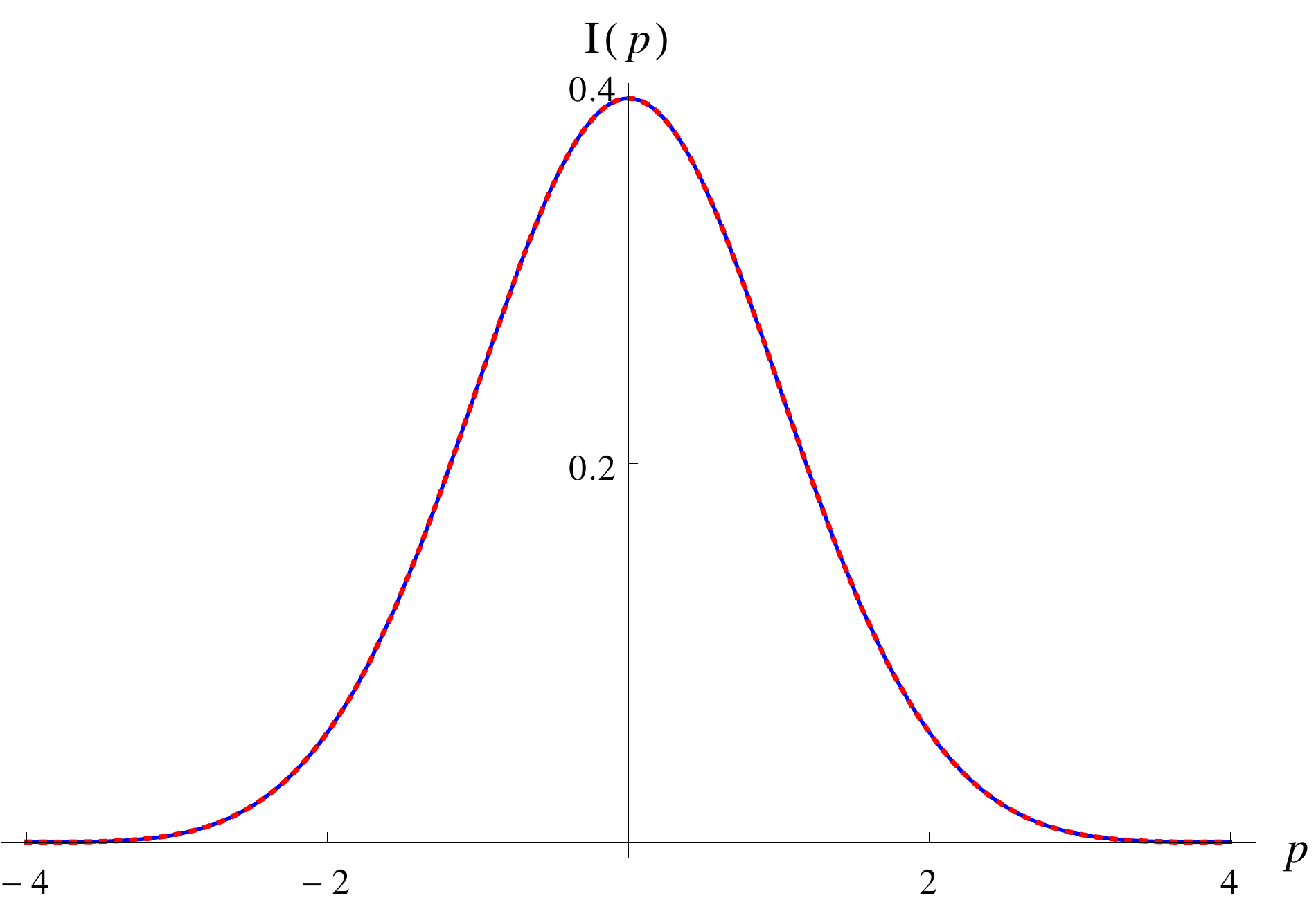}
	\caption{Momentum distribution $I(p)$ for the ground state using Eqs. (19,20) (solid line). Dashed line is due to numerical integration (2). Here $2m=1=\hbar^2$, $V_0=10$, $a=2$ and $\beta=1.1862$, see Eq. (11). The solid and dashed lines have merged.}
\end{figure}
\begin{figure}
	\centering
	\includegraphics[width=8 cm,height=5 cm]{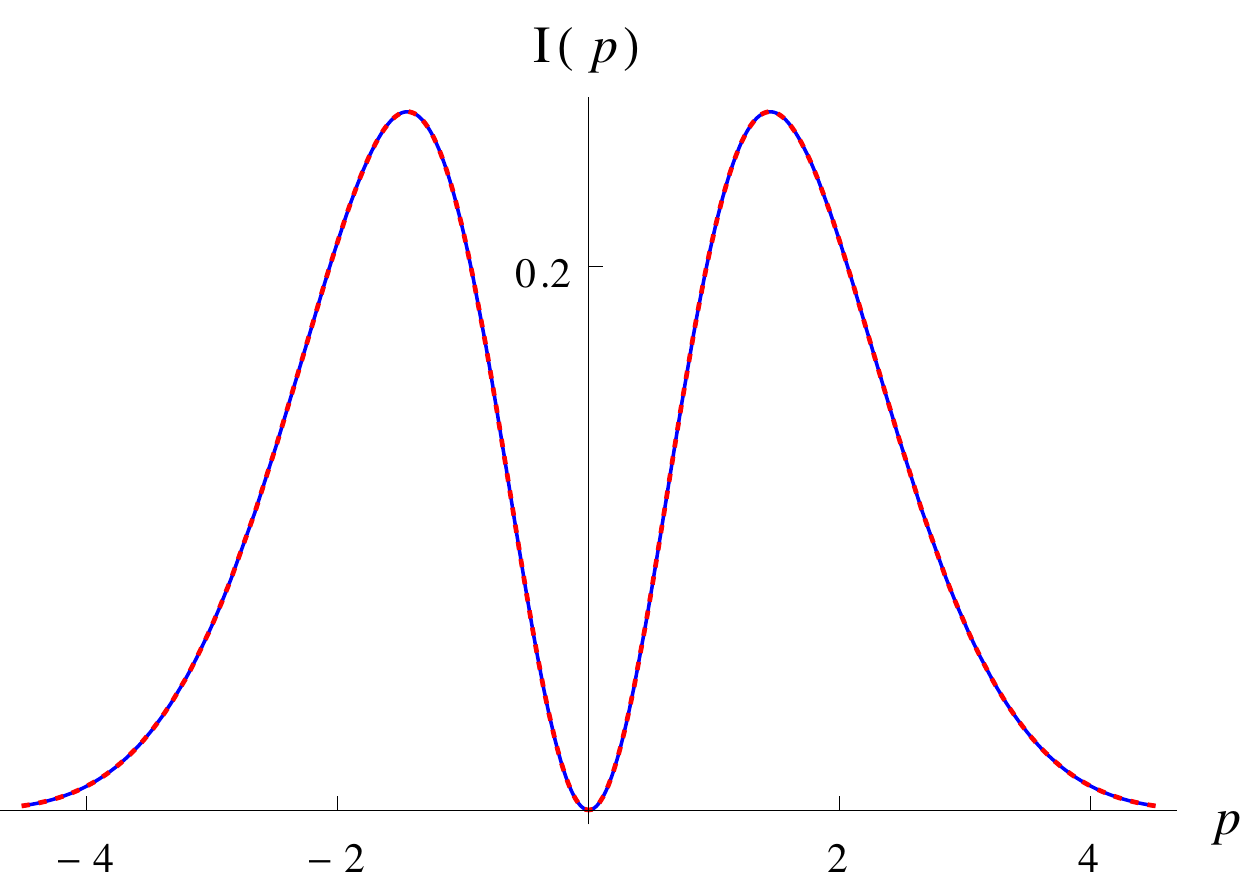}
	\caption{The same as in Fig. 1, for the first excited state. Here $\beta=2.3185$, see Eq. (13). Analytic form used here are in Eqs. (22,23).}
\end{figure}
\begin{figure}
	\centering
	\includegraphics[width=8 cm,height=5 cm]{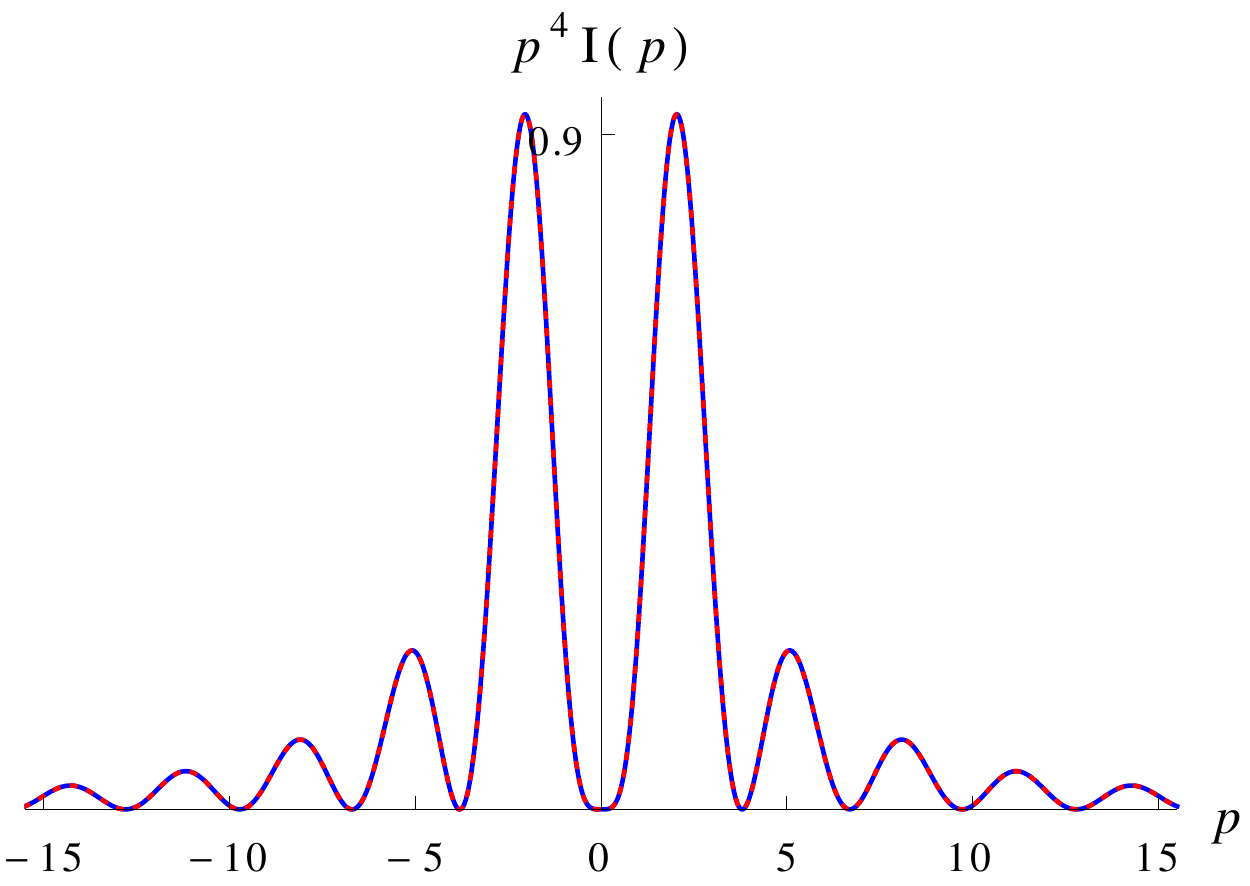}
	\caption{$p^4I(p)$ for the ground state of the FSW as in Fig. 1.}
\end{figure}
\begin{figure}
	\centering
	\includegraphics[width=8 cm,height=5 cm]{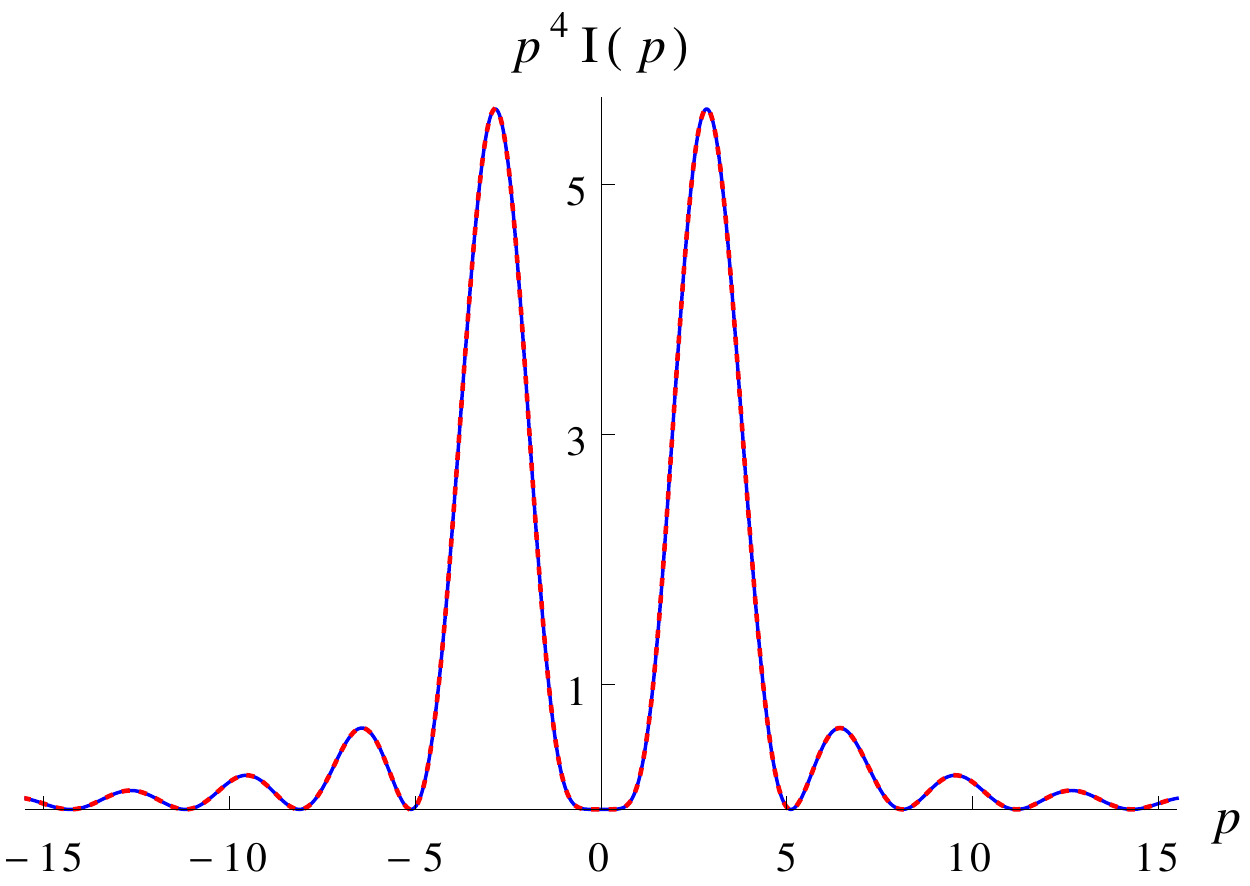}
	\caption{$p^4I(p)$ for the first excited state of the FSW as in Fig. 2}
\end{figure}

\begin{figure}
	\centering
	\includegraphics[width=8 cm,height=5 cm]{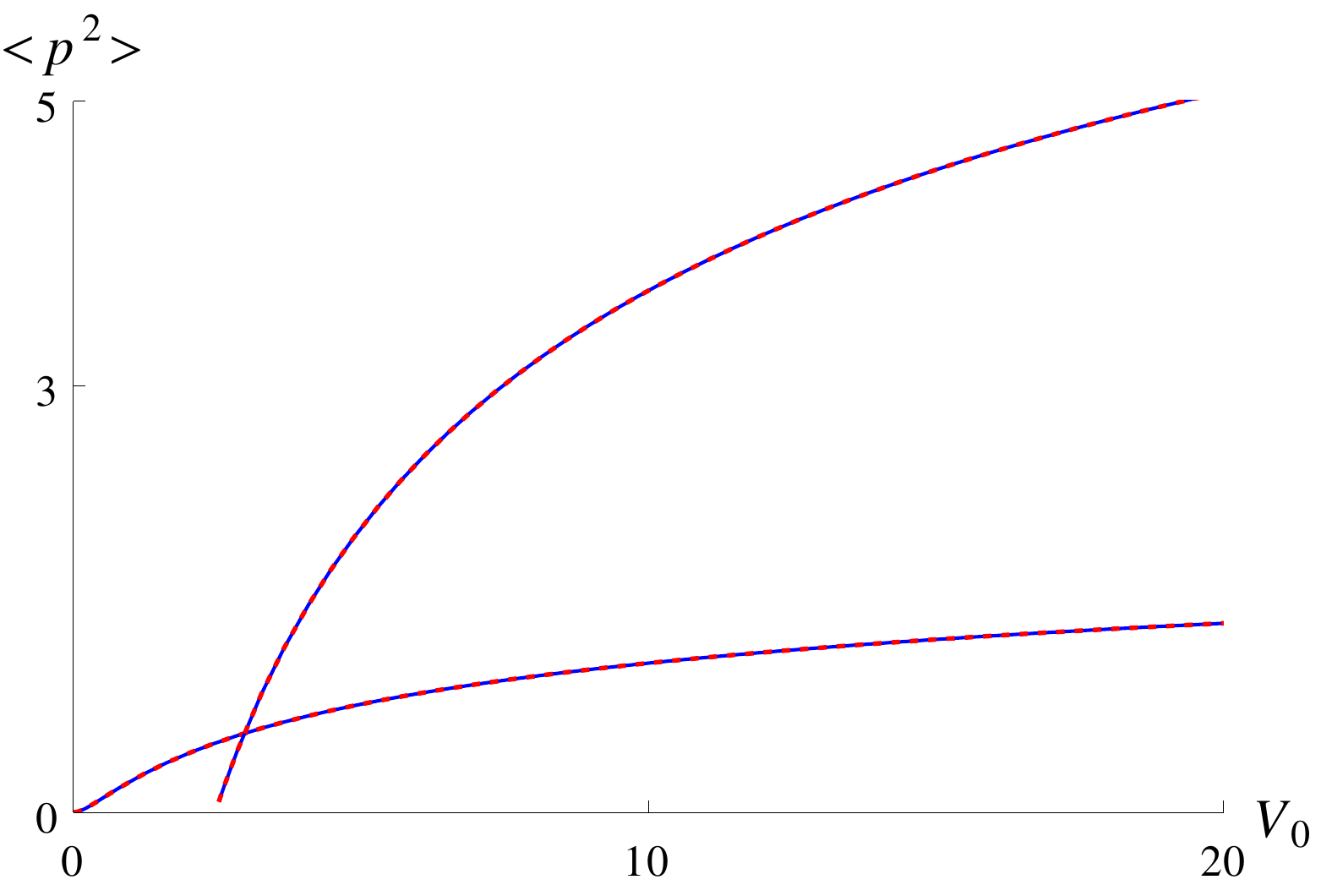}
	\caption{Expectation values of $p^2$ calculated from Eq.(15) (solid line) and from momentum space distribution $I(p)=|\phi(p)|^2$ numerically (dashed line) using Eqs. (19,20). The parameter $a$ is fixed as 2 and the well-depth $V_0$ is varied. Lower curve is for the ground state  and the upper one is for the first excited state.  Notice that dashed  lines and solid lines match very well, we have taken $p\in[-100,100]$.}
\end{figure}
\begin{figure}
	\centering
	\includegraphics[width=8 cm,height=5 cm]{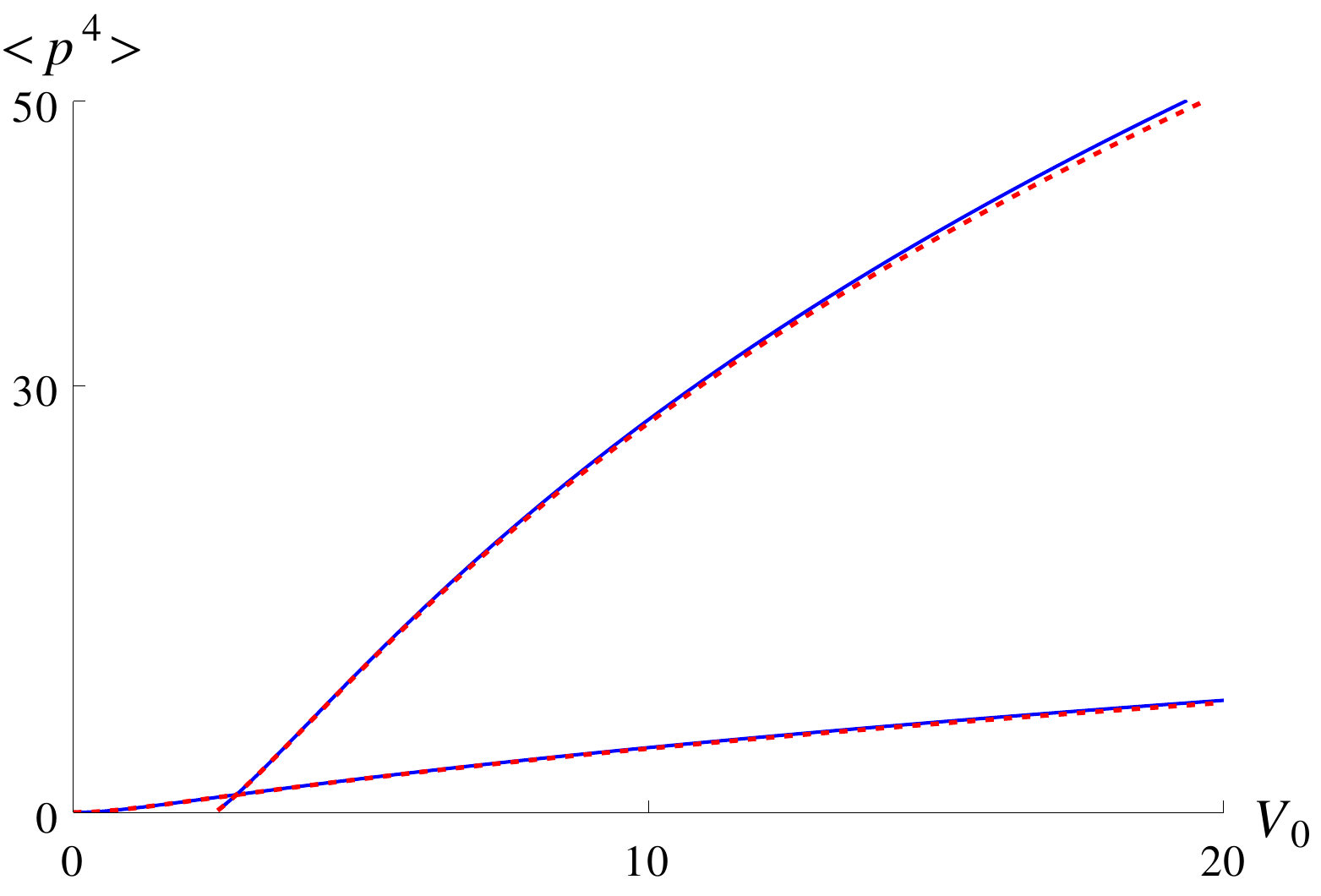}
	\caption{The same as in Fig. 5 for $<\!p^4\!>$ to  using Eq. (17) (solid line) and due to numerical integration (dashed line) using Eqs. (22,23). Here the dashed  and the solid lines agree fairly when we take $p\in[-100,100]$, this agreement can be improved easily by extending this domain.}
\end{figure}
For odd parity states (6) of FSW (2), we get
\begin{multline}
\hspace*{-0.4cm}\phi^{in}_o(p){=}\frac{2i A_o [\beta \cos d \sin(za/2){-}z \sin d \cos (za/2)]}{\sqrt{2 \pi \hbar} (\beta^2{-}z^2)} \\ 
\hspace*{-0.38cm} \phi^{out}_o(p){=}\frac{-2i A_o[\alpha \sin (za/2){+} z \cos(za/2)]\sin d}{\sqrt{2\pi \hbar}(z^2{+}\alpha^2)}, z{=}\frac{p}{\hbar}.
\end{multline}
Once again $\phi^{in}_o(p)$ and $\phi^{out}_o(p)$ seem to vary asymptotically  as $p^{-1}$, but when we use 
the eigenvalue condition (13), we get
\begin{eqnarray}
\hspace*{-.15 cm} \phi_o(p){=} \frac{-2i A_o \gamma^2 \sin d~[\alpha \sin(za/2)+z  \cos(za/2)]} {\sqrt{2\pi \hbar} (z^2+\alpha^2)(z^2-\beta^2)}   {\sim} \frac{1}{p^3}.
\end{eqnarray}
$\phi_o(p)$ in (22) differs from Eq. A.1 of [8] in details. In [8] $\phi_o(\hbar\beta)$ diverges but in (22) it becomes $0/0$, wherein the  limit can be found by using L'Hospital's rule as
\begin{equation}
\hspace*{-.15 cm}\lim_{z\rightarrow \beta} \frac{[\alpha \sin(\frac{za}{2})+z \cos(\frac{za}{2})]}{(z^2-\beta^2)} {\rightarrow} 
\frac{(\alpha a{+}2)\cos d{-}2d \sin d}{4\beta}
\end{equation}
\section{V. Discussion and Conclusion}
We take $2m{=}1{=}\hbar^2$, $V_0{=}10$, $a{=}2$ in arbitrary units to present the momentum distribution $I(p)=|\phi(p)|^2$ for ground state and the first excited state of FSW potential in Figs. 1 and 2. In Figs. 3 and 4 we plot $p^4 I(p)$ to show that these distributions do converge asymptotically but not without oscillations. In these Figs. 1-4, the solid lines are due to the forms derived (19, 22) by us. The agreement between  solid and dashed lines (numerical integration) testifies to the correctness of our analytic forms. We have also checked our analytic forms  for several sets of values of $V_0$ and $a$ by numerical integrations. We remark that our analytic forms (14, 19, 22) suggest various corrections to results given in Ref. [8]. Even the verification that the momentum space eigenfunctions  Eqs. (19, 22) eventually satisfy the integral equation (3) has been done numerically by us. The analytic expressions (20, 23) of limits derived by us are new and they help in evaluating various $p$-integrals which become improper [7]. 

Our Eqs. (15) and (17) respectively for $<\!p^2\!>$ (solid line in Fig. 5) and $<\!p^4\!>$ ( solid line in Fig. 6) for FSW are new which are derived in position space by circumventing the discontinuity of second and higher derivatives of $\psi(x)$ at the end points. We also find  $<\!p^2\!>$ and $<\!p^4\!>$ in the momentum space  using (19, 22) by doing $p$-integrals numerically (dashed lines in Figs. 5 and 6), these integrals are improper [7] but convergent. By fixing $a=2$ and varying $V_0$, in Fig. 5 and 6, we present $<\!p^2\!>$ and $<\!p^4\!>$, respectively; for the ground state (lower line) and the first excited state (upper line).  In Fig. 5 and 6, we have taken $p \in [-100,100]$ to display a fair agreement between  dashed and solid lines. These domains can be extended to improve the agreement between the two easily.

Finally, we conclude that the  present paper provides correct expressions (19, 22) of momentum space eigenfunctions $\phi(p)$ of the finite square well potential which seem to be unavailable otherwise. The expressions (15, 17) for $<\!p^2\!>$and $<\!p^4\!>$ obtained in a simple way in position space are new. This revisit to finite square well potential brings out the need to study the fall-off of  $\phi(p)$ in other potential wells which are defined piece-wise. Such models are finite triangular well: $(V(|x| \le a)=-V_0(1-|x/a|)$ and finite parabolic well ($V(|x|
\le a)=-V_0(1-(x/a)^2)$), where $V(|x|>a)=0$. In these models it is the third derivative of $\psi(x)$ which would be discontinuous at $x=\pm a$. This paper is instructive wherein subtle mathematical and procedural differences between two (position and momentum) representations of wavefunctions have been brought out.

We would like to remark that the analytic verification that $\phi(p)$ given in Eqs. (19) and (22) satisfy the Schr{\"o}dinger equation in momentum space (3) would be educative which is due. To the best of our knowledge, the analytic solution of the integral equation (3) for FSW to obtain Eqs.(19) and (22) and the analytic expressions for $<\!p^2\!>$ and $<\!p^4\!>$ thereof remain due to be done next.


\end{document}